# Unconventional critical behaviors at the magnetic phase transition of $Co_3Sn_2S_2$ kagomé ferromagnet

Mohamed A. Kassem[1,2,*], Yoshikazu Tabata[1], Takeshi Waki[1], Hiroyuki Nakamura[1]

[1] Department of Materials Science and Engineering, Kyoto University, Kyoto 606-8501, Japan,

[2] Department of Physics, Faculty of Science, Assiut University, 71516 Assiut, Egypt.

**Abstract**

$Co_3Sn_2S_2$ has generated a growing interest as a rare example of the highly uniaxial anisotropic kagomé ferromagnet showing a combination of frustrated-lattice magnetism and topology. Recently, via precise measurements of the magnetization and AC susceptibility we have found a low-field anomalous magnetic phase (A-phase) with very slow spin dynamics that appears just below the Curie temperature ($T_C$). The A-phase hosts high-density domain bubbles after cooling through $T_C$ as revealed in a previous *in-situ* Lorentz-TEM study. Here, we present further signatures of the anomalous magnetic transition (MT) at $T_C$ revealed by a study of the critical behaviors of the magnetization and magnetocaloric effect using a high-quality single crystal. Analyses of numerous magnetization isotherms around $T_C$ ($\simeq$ 177 K) using different approaches (the modified Arrot plot, Kouvel-Fisher method and magnetocaloric effect) result in consistent critical exponents that do not satisfy the theoretical predictions of standard second-order-MT models. Scaling analyses for the magnetization, magnetic entropy change and field-exponent of the magnetic entropy change, all consistently show low-field deviations below $T_C$ from the universal curves. Our results reveal that the MT of $Co_3Sn_2S_2$ can not be explained as a conventional second-order type and suggest an anomalous magnetic state below $T_C$.

## 1. Introduction

The yield of multi-$q$ states in magnetic systems with competing interactions has become emerging frontier in condensed matter physics. Many topological phenomena, related to the nontrivial $\pi$ Berry phase captured by electrons circulating around the integer-Chern number points, such as the multiferriocity and large anomalous Hall and Nernest effects have been observed in these systems[1–6]. A recently proposed model platform along these lines is the kagomé lattice of magnetic atoms[7]. Exotic states of topological nature such as magnetic skyrmions and Dirac fermionic states have been observed in kagomé magnets[8–11]. One kagomé ferromagnet that has received a lot of interest is $Co_3Sn_2S_2$. The compound exhibits interesting magnetic properties originating from its magnetism of quasi-two-dimensional nature with strong magnetic anisotropy[12–14]. Furthermore, it has attracted attention as a model magnetic Weyl semimetal with flat bands[9,10,15].

$Co_3Sn_2S_2$ crystalizes in the shandite miniral, $Ni_3Pb_2S_2$, structure (R-3m symmetry with Co at $9e(1/2,0,0)$, S at $6c(0,0,z)$ and Sn atoms occupy $3b(0,0,1/2)$ and $3a(0,0,0)$ sites). that has been presented in many configurations [9,12,16–19]. As shown in Fig.1(a), it can simply be considered as parallel metallic layers of $Co_3$-Sn with Co atoms of magnetic moments in the $c$-axis direction are arranged on a kagomé sublattice, Fig. 1(b), at a Co-Co distance of about 2.7 Å. The layers are separated by Sn-$S_2$ blocks with an interlayer distance of 4.4 Å.

We have previously measured the temperature-dependent static magnetization and AC susceptibility and reported interesting features using single crystal grown by a flux method[20]. The occurrence of an anomalous magnetic phase (A-phase) at fields below about 0.06 T in the vicinity of the Curie temperature ($T_C \approx 173$ K), large splits between the field-cooled (FC) and zero-field-cooled (ZFC) magnetizations only in the low-field region, and extremely slow spin dynamics with relaxation times longer than 10 s have been observed. Quite recently we have studied the magnetic domain evolution in the A-phase in an *in-situ* Lorentz microscopic study[21]. The domain structure is typical for anisotropy-controlled material that strongly depends, as well as the domain wall (DW) motion, on the temperature, its change rate and process (heating or cooling) and the applied magnetic field. Inactive magnetic domains due to strong pinning in the low-temperature range, which should be attributed to the strong uniaxial anisotropy in this compound, is activated with increasing temperature, as a result of thermal depinning, or with increasing the applied field. The thermal activation of a domain wall motion approximately starts around the lower-temperature boundary of the A-phase, denoted as $T_A$ in Ref. [20]. Interestingly, a spontaneous domain reorganization which is associated with DW creep has been observed in the A-phase even in the absence of an external field. While cooling the crystal through $T_C$, high-density mixture of rounded magnetic bubbles and chain-like domains in zero field and only bubbles at 0.001 T became visible just below $T_C$. In agreement with the results of the frequency-dependent AC susceptibility, the DWs were nearly static on a time scale of 10 s below 172 K. We attributed the occurrence of the A-phase to the formation of high-density bubble domains, possibly of nontrivial spin textures, and their spontaneous reorganization in the low-field region and below $T_C$[21]. Recent muon spin-rotation ($\mu$SR) results have presented the possibility of an emergent competitive in-plane antiferromagnetic (AFM) order that tune the Berry-curvature anomalous Hall conductivity in the temperature range of the A-phase[22].

One way to investigate for the nature of a magnetic ground state is to study the magnetic transition (MT) by analysis for the critical behavior and/or corresponding magnetocaloric effect (MCE)[23,24]. Although the low-field A-phase of $Co_3Sn_2S_2$ is controversial, the MT can be studied as a standard second-order MT when viewed from higher fields, for example above the saturating field. Here, we study the critical behaviors of the magnetization and MCE using a high-quality single crystal of $Co_3Sn_2S_2$, which present further signatures of the MT at $T_C$. In consistence with the reported low-field anomalous magnetic state[20], critical scaling analyses of the magnetization and magnetic entropy changes show consistent deviations from the second-order MT universal curves, appear below $T_C$ only at low magnetic fields ($\leq 0.1$ T).

## 2. Experimental procedures

A single crystal of $Co_3Sn_2S_2$ used in this study was grown in a tipped glassy-carbon crucible of 1 cm inner diameter that was sealed in an evacuated quartz ampoule as in a previously described modified Bridgman method[25]. About 10-g crystal has been removed from the crucible and easily cleaved in the (001) plane into two equal parts, one of them is shown in the inset of Fig. 1(c). Crushed parts were investigated by powder x-ray diffraction (XRD) and the stoichiometric chemical composition was confirmed by the wave-length dispersive x-ray spectroscopy (SEM–WDX). The single crystallinity and orientations have been identified by Laue x-ray spectroscopy that revealed the flat shiny surface as the (001) plane. Both crystals used in the current study and that recently used in a Lorentz TEM study[21] were selected from the same batch. Magnetization measurements were performed using a Quantum Design, Magnetic Property Measurement System (MPMS). The temperature dependence of the magnetization has been measured at different fields from 0.5 to 100 mT following the same protocols that we previously used for a flux-grown crystal[20]. The magnetic entropy change and critical exponents were estimated from magnetization isotherms measured at temperatures from 90 to 200 K with a step of 2 K (reduced around $T_C$ to 0.5 K) after zero-field cooling from room temperature and while decreasing the applied magnetic field from 5 to 0 T.

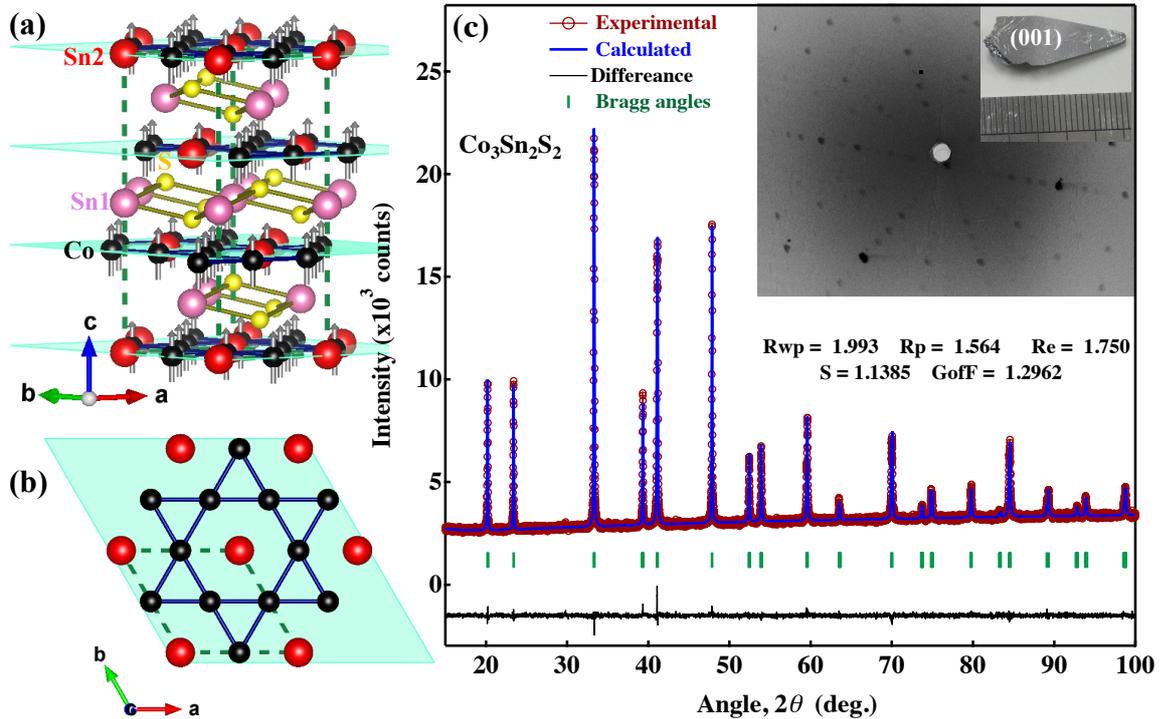

**Figure 1: Structural properties and crystal quality**: **(a)** the crystal structure of $Co_3Sn_2S_2$ showing the metallic $Co_3$-Sn layers separated by Sn-$S_2$ blocks along the *c*-axis with **(b)** one layer contains a Co-kagomé sublattice projected to the (001) plane. The magnetic moments are shown along the easy *c*-axis. **(c)** experimental (symbols) and Rietveld-calculated (thick solid) powder XRD patterns for crushed parts of $Co_3Sn_2S_2$ crystal grown by a modified Birdman method (see text), shown in the inset on a mm scale. Laue x-ray patterns on the shown cleavage plane (001) with the six-fold symmetry of the hexagon structure are presented in the inset background of (c).

## 3. Results

### 3.1. Crystal quality and structure

The high quality of the synthesized large single crystal of $Co_3Sn_2S_2$ is seen in the cleaved shiny surface and clearly indicated in its Laue patterns (inset of Fig. 1(c)) showing the bulk-structure six-fold symmetry of the cleavage plane (001). Experimental powder XRD pattern of crushed parts of the synthesized crystal is shown with calculated pattern in the main panel of Fig. 1(c). No lines of extra phases rather than those of the shandite structure appear. Refinement of the experimental structure parameters by a Rietveld analysis results in lattice constants of $a$ = 5.3689(1) Å (intralayer Co-Co distance of 2.684 Å) and $c$ = 13.1786(3) Å (interlayer distance of 4.392 Å) and sulfur atomic position $z$ = 0.2146(1). A piece of the shown single crystal in the inset of Fig. 1(c) has been used for all measurements of the magnetization presented here below, the temperature dependence at different applied magnetic fields and the isothermal magnetization curves, $M$ vs. $H$, at different temperatures.

### 3.2. Temperature dependence of magnetization

The temperature dependence of magnetization measured at low fields of 0.5 and 5 mT after ZFC and FC processes is presented in Figs. 2(a) and 2(b) for fields applied along and perpendicular to the $c$ axis, respectively. The low-field $M(T, H)$ with both $H // c$ and $H \perp c$ shows a sudden increase at the Curie temperature $T_C$, which corresponds the reported FM transition[20,26,27]. While the observed slightly higher $T_C$ = 177 K, comparted to $T_C$ = 174 K (±1 K) exhibited by flux-grown crystals[12,18–20,27], indicates a sample dependence[20], common features of the low-field $M(T, H)$ in the present and previous studies are found: the sudden increase of $M(T, H)$ at $T_C$, the large split of the ZFC and FC magnetization curves, humps in the FC-$M(T, H)$ at temperatures $T_A$ and minima in-between $T_C$ and $T_A$ in the ZFC-$M(T, H)$. Other similarities are also found in the higher field $M(T, H)$ as shown in Figs. 2(c) and (d), namely, the anomalies are suppressed with increasing the magnetic field and disappear at and above fields compared to the saturating magnetic field, which is too high to observe when applying field in the *ab*-plane due to the high magnetic anisotropy[12,13]. All observed anomalies indicate the reported low-field anomalous magnetic state close to $T_C$ (*A*-phase) of $Co_3Sn_2S_2$.

For approaching accurate critical exponents and hence study the intrinsic scaling behaviors of the magnetization and corresponding magnetic entropy change, the influence of demagnetizing field has to be ruled out[28]. To estimate the sample-shape-dependent demagnetization factor, $D$, initial magnetization curves have been measured at different temperatures below $T_C$, as shown in the inset of Fig. 2(a). The initial isothermal curves at all temperatures almost exhibit identical slopes before the magnetization saturation, where a zero-internal-field condition $H_{int} = H - DM = 0$ is satisfied. The value of $D$ can be estimated as the inverse of slope, $D = H_c/M_s$, where $H_c$ and $M_s$ are the saturating field and magnetization, respectively. A value of $D = 0.331$ is obtained for our used sample of an almost cubic shape. The magnetization and other physical quantities, such as the magnetic entropy change and its field-exponent, here below are presented as functions of $H_{int}$.

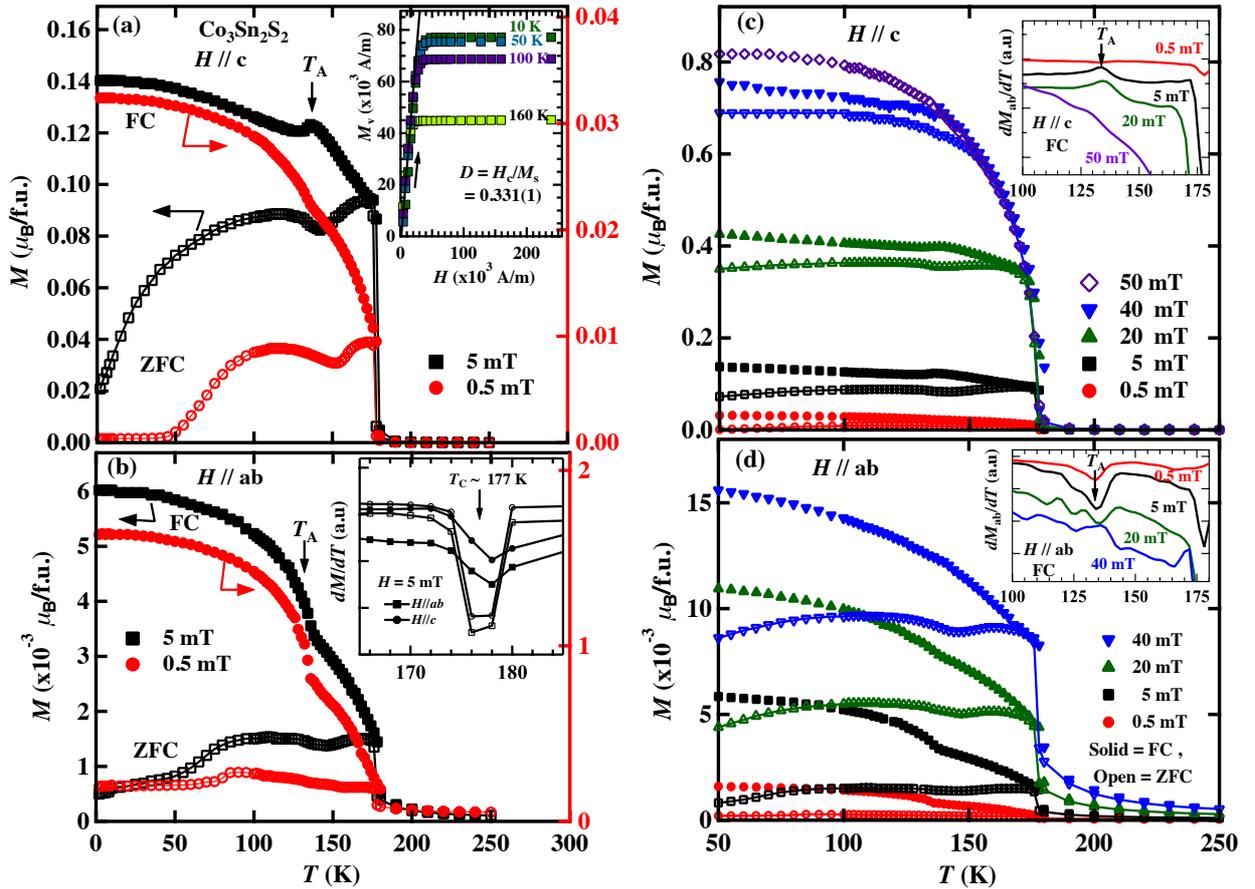

**Figure 2: Temperature dependence of magnetization *M(T, H)* of $Co_3Sn_2S_2$:** *M(T, H)* measured at different fields applied **(a)**, **(c)** along and **(b)**, **(d)** perpendicular to the *c* axis after field-cooling (closed) and zero-field cooling (open) processes and measured. Vertical arrows indicate the anomalous phase boundary, $T_A$. The inset of (a) shows the initial magnetization curves (volume magnetization against external field) at different temperatures and insets of (b-d) show the temperature dependence of the magnetization derivative, d*M(T, H)*/d*T*, at indicated fields and their directions after a field-cooled process.

### 3.3. Critical scaling analysis of magnetization

To study the critical behavior of $Co_3Sn_2S_2$, magnetization isotherms have been measured with applying fields in the easy $c$-axis. The critical scaling analysis of magnetization is described in detail in the Appendix A.

Figure 3(a) shows the isothermal magnetization as a function of the internal field, $M(H_{int}, T)$, measured at temperatures from 90 K to 200 K with a temperature interval of 2 K. The critical $M(H_{int}, T_C)$ curve at the Curie-temperature $T_C = 177$ K is shown by circles. The $M(H_{int})$ curves at $T < T_C$ show a part of square shape that becomes almost linear for $T > T_C$ which is a typical behavior for a ferromagnet.

The well-known Arrot plots, $[M(H_{int})]^2$ vs. $\mu_0 H_{int}/M$, are shown in Fig. 3(b). The standard Arrot plot is based on the equation of state with the mean field criticality, Eq. (A2) in Appx. A, that predicts parallel straight lines of $[M(H_{int})]^2$ vs. $\mu_0 H_{int}/M$. In consistence with previous observations[12,14], the Arrot plots of $Co_3Sn_2S_2$ exhibit linearity far above and below $T_C$, whereas, plots around $T_C$ are concave curves to higher magnetization particularly in the low field region. This concaved behavior of $[M(H_{int})]^2$ vs. $\mu_0 H_{int}/M$ in the vicinity of $T_C$ indicates a non-mean-field criticality of the MT of the Co-shandite.

To approach the values of critical exponents of $Co_3Sn_2S_2$, we have first approximated the spontaneous magnetization, $M_s(T)$, below $T_C$ and the initial inverse susceptibility, $\chi_0^{-1}(T)$, above $T_C$ by extrapolating the high-field linear Arrot plot portions to $M(H_{int}) = 0$ and $\mu_0 H_{int}/M = 0$ for $T < T_C$ and $T > T_C$ data, respectively. The temperature dependences of the $M_s(T)$ and $\chi_0^{-1}(T)$ estimated from the standard Arrot plots are presented in the inset of Fig. 3(b). Fittings to Eqs. (A4) and (A5) in Appx. A, shown by solid lines, for $M_s(T)$ and $\chi_0^{-1}(T)$ result in critical exponents of $\beta = 0.358(9)$ and $\gamma = 1.054(24)$ as well as approximated $T_C = 178.8$ and $179.25$ K, respectively. Introducing these values of $\beta$ and $\gamma$ to the Widom scaling relation results in $\delta = 1 + \gamma/\beta = 3.944(99)$. The inconsistently estimated critical exponents from the standard Arrot plot, namely the non-mean-field exponents estimated from the mean-field equation of state presented in table 1, indicates violation from the mean-field approximation in $Co_3Sn_2S_2$, as well as the concaved behavior of the Arrot plots does.

Hence, we have examined a further analysis using the Arrot-Noaks equation, Eq. (A3), with consideration of a non-mean-field criticality, i.e., the modified Arrot plot, $[M(H_{int})]^{1/\beta}$ vs. $(H_{int}/M)^{1/\gamma}$. If appropriate critical exponents $\beta$ and $\gamma$ are adopted, the experimental data exhibits parallel straight lines and the temperature dependences of the estimated $M_s(T)$ and $\chi_0^{-1}(T)$ would be correctly described by using the adopted exponents. Practically, the values of $\beta = 0.358$ and $\gamma = 1.054$, obtained from the standard Arrot plot (Fig. 3(b)) have been used initially to establish a modified Arrot plot. Again, we estimated $M_s(T)$ and $\chi_0^{-1}(T)$, from the established modified Arrot plot data and obtained the critical exponents $\beta$ and $\gamma$ by fitting these $M_s(T)$ and $\chi_0^{-1}(T)$ to Eqs. (A4) and (A5). The obtained exponents were used for remaking the modified Arrot plot and the procedure was iterated until parallel straight lines with unchanged critical exponents[29] have been obtained. Critical exponents and MT temperature, $\beta$ = 0.343(2), $T_C$ = 177.14 K and $\gamma$ = 1. 116(15), $T_C$ = 177.13 K are obtained from data at $T < T_C$ and $T > T_C$, respectively. Figure 3(c) shows the modified Arrot plots $[M(H_{int})]^{1/\beta}$ vs. $(H_{int}/M)^{1/\gamma}$, with the finally obtained critical exponents in this procedure. Nicely linear relation between $[M(H_{int})]^{1/\beta}$ and $(H_{int}/M)^{1/\gamma}$ is found, indicating a valid set of the critical exponents are estimated.

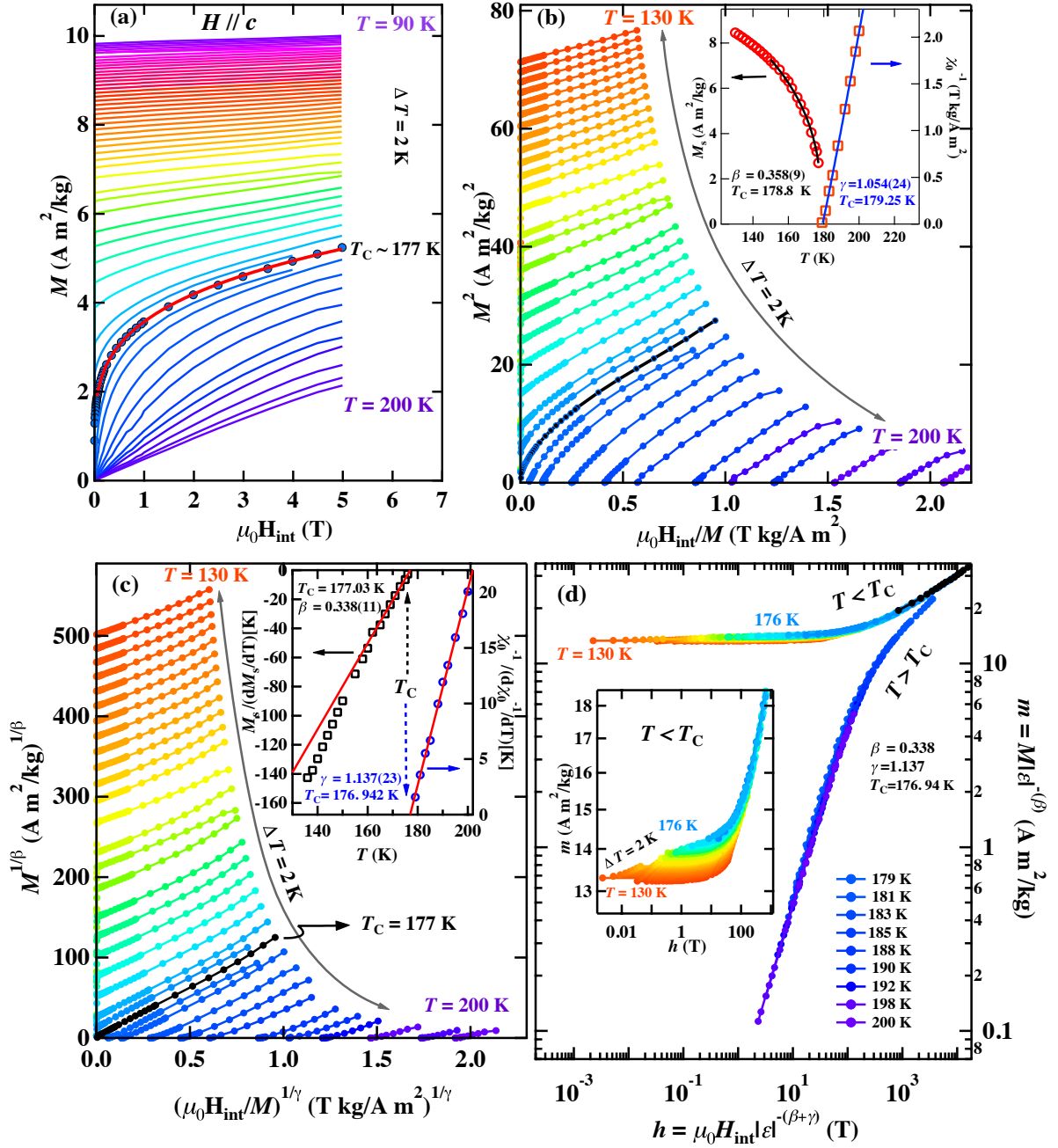

**Figure 3: Critical phenomenon of $Co_3Sn_2S_2$:** Easy-axis magnetization isotherms measured at temperatures blow and above the shown Curie-temperature, $T_C$, while decreasing field from 5 T after a ZFC process, presented in the form of **(a)** $M(T,H)$ vs. $H$, **(b)** Arrot plots, $M^2(T,H)$ vs. $H/M$, and **(c)** modified Arrot plots (Arrot-Noaks form). The insets in (b) and (c) show the $T$-dependence of the saturated magnetization $M_s$ and inverse susceptibility $\chi^{-1}(T,H)$ estimated by extrapolations of the Alrrot plots at $T < T_C$ and $T > T_C$, respectively, at high-fields as shown on linear scales in (b) and at the whole range of $H$ after back and forth iterations in a Kouvel-Fisher form in the inset of (c). **(d)** Scaling plot of the ($M$-$T$-$H$) data at $T < T_C$ and $T > T_C$, based on Eqs. (A9) and (A10) in Appex. A, using the estimated critical exponents.

An alternative procedure to estimate the values of $\beta$, $\gamma$ and $T_C$ is well-known as the Kouvel-Fisher method[30]. The inset of Fig. 3(c) shows the linear relations of $M_S(T)[dM_S(T)/dT]^{-1}$ vs. $T$ and $\chi_0^{-1}(T)[d\chi_0^{-1}(T)/dT]^{-1}$ vs. $T$ of $Co_3Sn_2S_2$. Linear fittings to Eqs. (A7) and (A8) result in values of $\beta$ = 0.338(11), $T_C$ = 177.03 K and $\gamma$ = 1. 137(23), $T_C$ = 176.94 K from data at $T < T_C$ and $T > T_C$, respectively, which are close to values from direct power-function fittings to Eqs. (A4) and (A5). Introducing the $\beta$ and $\gamma$ values obtained by the above two methods to the Widom relation results in $\delta$ = 4.254(48) and 4.364(129), respectively. Further, the critical exponent $\delta$ can be obtained by fitting the experimental magnetization isotherm data measured at $T \simeq T_C$ to Eq. (A6). The fitting results is shown by the solid line at $T$ = 177 K in Fig. 3(a) with $\delta$ = 4.441(19), which is a slightly higher than the values extracted from the Arrot-Noaks plot by Widom relation, that may be due to the slight difference in the assumed $T_C$. The correct approximation to the critical exponents $\beta$, $\gamma$ and $\delta$ is confirmed by the self-consistency among various procedures of estimation which are summarized in table 1.

An analysis of the magnetization data based on the scaling hypothesis to a universal curve for second-order MTs, described in Appx. A, can help us to approximate the critical regime of $Co_3Sn_2S_2$ and may imply features of its controversial MT. Although the MT of $Co_3Sn_2S_2$ has been recently proposed as a second-order type[14], the abruptly increased magnetization at $T_C$ and the anomalies observed in $M(T)$ below $T_C$ with applying low magnetic fields, Fig. 2, all indicate a possible unconventional MT. Investigation of the critical behavior of $Co_3Sn_2S_2$ carefully performed close to the MT, i.e., at $\mu_0 H_{int} \to 0$, would be useful.

Figure 3(d) shows the scaling plot, $M(H_{int}, \varepsilon)|\varepsilon|^{-\beta}$ vs. $\mu_0 H_{int}|\varepsilon|^{-(\beta+\gamma)}$, where $\varepsilon = (T - T_C)/T_C$ is the reduced temperature, for all the $M$-$T$-$H_{int}$ data of $Co_3Sn_2S_2$ shown on a log-log scale for clarity. The scaled plot is generated based on the asymptotic equation of state in the critical regime, Eqs. (A9) and (A3), by using the critical exponents estimated by the Kouvel-Fisher method from the modified Arrot plot. As predicted by the scaling hypothesis, it can be seen that most of the magnetization data fall onto two universal curves, one for $T > T_C$ and another for $T < T_C$ data. However, close inspection to the $T < T_C$ data, as shown in the inset of Fig. 3(d), shows that the well scaled data to the universal behavior occurs only in the high fields region while the low-field data show deviations from the universal curve. The deviation in this plot occurs for the temperature-curves with $T < T_C$ down to about 140 K at crossover fields of about 0.07 - 0.2 T, which indicates an unconventional MT to a low-field anomalous magnetic state, as will be further discussed later below.

### 3.4. Critical scaling study of the magnetocaloric effect

As well as its technological interest[31], the magnetocaloric effect (MCE) acts as an effective method to study the magnetic phase transition and even is useful to probe the nature of a magnetic ground state[24,32–37]. The magnetic entropy change $\Delta S_M(T, H_{int}) = S(T, H_{int}) - S(T, 0)$ exhibits the critical scaling behavior[38], as well as the magnetization does, and also its sign indicates how the system is ordered by magnetic field with respect to its zero-field magnetic preferred configuration, for instance, the ferromagnetic and antiferromagnetic states correspond to $\Delta S_M < 0$ and $\Delta S_M > 0$, respectively. We have studied the MCE of $Co_3Sn_2S_2$ by estimating $\Delta S_M$ based on the thermodynamic Maxwell relation after corrections for the demagnetizing field and by using the non-zero internal-field magnetization data. The derivation of $\Delta S_M$ and its critical scaling phenomena are described in Appx. B.

Figure 4(a) shows the temperature dependence of $\Delta S_M(H_{int}, T)$ of $Co_3Sn_2S_2$ estimated using $H_{int}$ and corresponding $M(H_{int}, T)$ data measured at different applied fields from 0.02 up to 5 T in a temperature range from 90 K up to 200 K. The negative values of the entropy change at all temperatures in the whole range of $H$ corresponds to the ferromagnetically ordered ground state of $Co_3Sn_2S_2$. The peak position of $\Delta S_M$ for different field scans occurs at about 177 K where the Curie temperature $T_C$ as the same as indicated by the critical scaling analyses of magnetization described in Sec. 3.3.

Similar to the critical isothermal $M(H_{int}, T_C)$, the absolute value of the magnetic entropy change at $T_C$, $|\Delta S_M^{peak}(H_{int}, T_C)|$; the peak width of $\Delta S_M(H_{int}, T)$ around $T_C$, $\delta T_{FWHM}(H_{int})$; and the

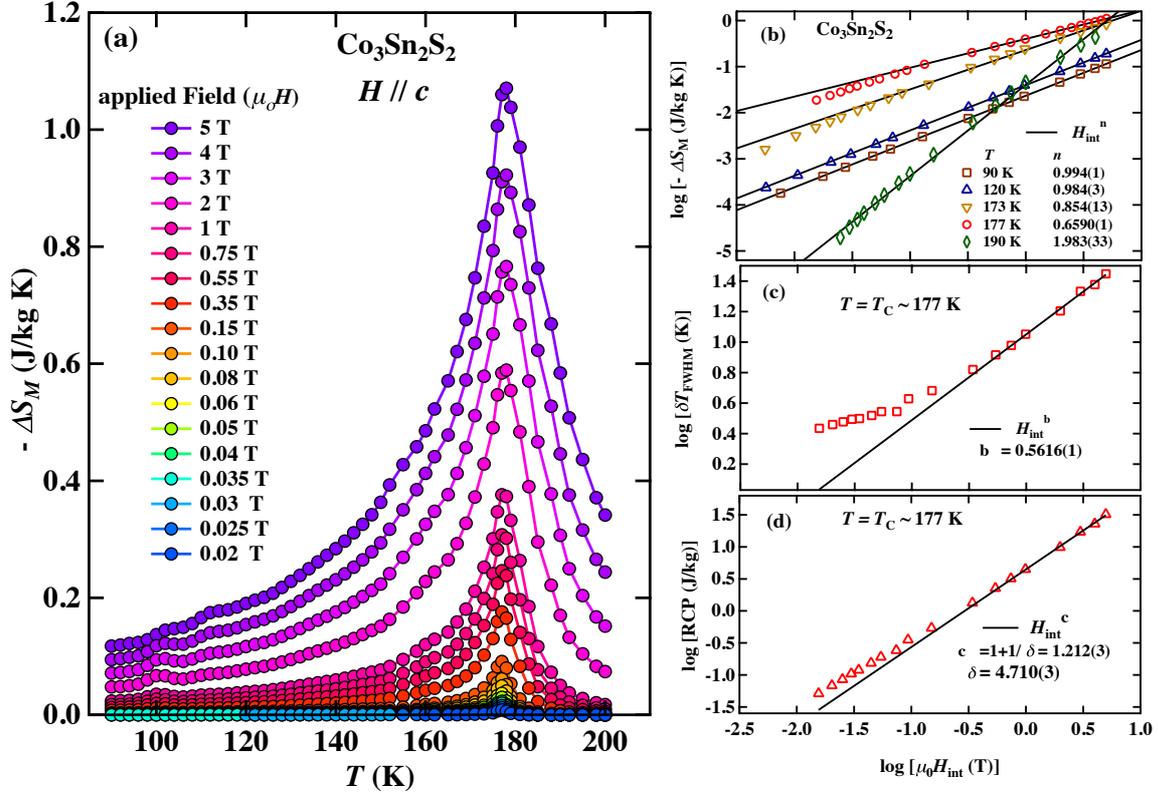

**Figure 4: Magnetocaloric effect of $Co_3Sn_2S_2$:** (a) Temperature dependence of the magnetic entropy change, $-\Delta S_M$, at low and high fields. The $H$ dependence of the estimated (b) $|\Delta S_M(H_{int}, T_C)|$, (c) its peak width $\delta T_{FWHM}$ (d) relative cooling power RCP = $|\Delta S_M^{peak}| \times \delta T_{FWHM}$ at indicated temperatures in log-log plots. The values of exponents $n$, $b$, $c$ and $\delta$ are estimated by fittings to Eqs. (B4) and shown by solid lines.

relative cooling power, $RCP = |\Delta S_M^{peak}(H_{int}, T_C)| \times \delta T_{FWHM}(H_{int})$, all increase with the magnetic field in power law behaviors[24], as Eqs. (B4).

Figures 4(b) - (d) show the internal-field dependences of the estimated critical $|\Delta S_M(H_{int}, T_C)|$, $\delta T_{FWHM}$ and $RCP$ presented as log-log plots. The three quantities estimated at $T_C$ exhibit linearity in the double-logarithmic plots, except for internal fields lower than about 0.09 T ($\log(\mu_0 H_{int}) \lesssim -1$). The MCE critical exponents are estimated by fittings Eqs. (B4) into the experimental data above 0.08 T, as $n(T_C) = 0.659$, $b = 0.5616$ and $c = 1.2123$. These MCE exponents are related with the critical exponents of magnetization as Eqs. (B5), and thus, the critical exponents $\beta = 0.3928$, $\gamma = 1.3878$ and $\delta = 4.7103$ are obtained, which are in consistence with values estimated from the critical scaling analyses of magnetization as presented in table 1.

Following the scaling procedure described in Appx. B, we have attempted the phenomenological universal scaling of the MCE[38]. We here rescale the $\Delta S_M(H_{int}, T)$ of Co$_3$Sn$_2$S$_2$ as $\left|\Delta S_M(H_{int},T)/\Delta S_M^{peak}(H_{int},T_C)\right|$ and plot against the reduced temperature, $\theta$, with appropriately selecting reference temperatures $T_{r1}(H_{int})$ and $T_{r2}(H_{int})$ based on its definition in Eqs. (B6) in Appx. B. The scaling plot is shown in Fig. 5(a) for all $\Delta S_M(H_{int}, T)$ data measured with field scans from different fields down to zero with the reference temperatures selected at the half maximum of $\Delta S_M(H_{int}, T)$, i.e., $T_{r2}(H_{int}) - T_{r1}(H_{int}) = \delta T_{FWHM}(H_{int})$. The normalized $\Delta S_M(H_{int}, T)$ curves measured below $T_C$ at high magnetic fields ($\mu_0 H \geq 0.1$ T, $\mu_0 H_{int} \gtrsim 0.08$ T) and those measured above $T_C$ at all fields are well-scaled to a universal curve previously reported in many experimental cases[24,34–36,38]. However, normalized $\Delta S_M(H_{int}, T)$ curves measured at lower fields show significant deviations from the universal curve just below $T_C$, as seen clearly in the inset of Fig. 5(a). The deviations are in consistence with deviations observed in the $M$-$T$-$H_{int}$ scaling plot in Fig.3(d) and the deviations from the linear-dependences in Figs. 4(b)-(d), in a further indication of the unconventional MT at $T_C$.

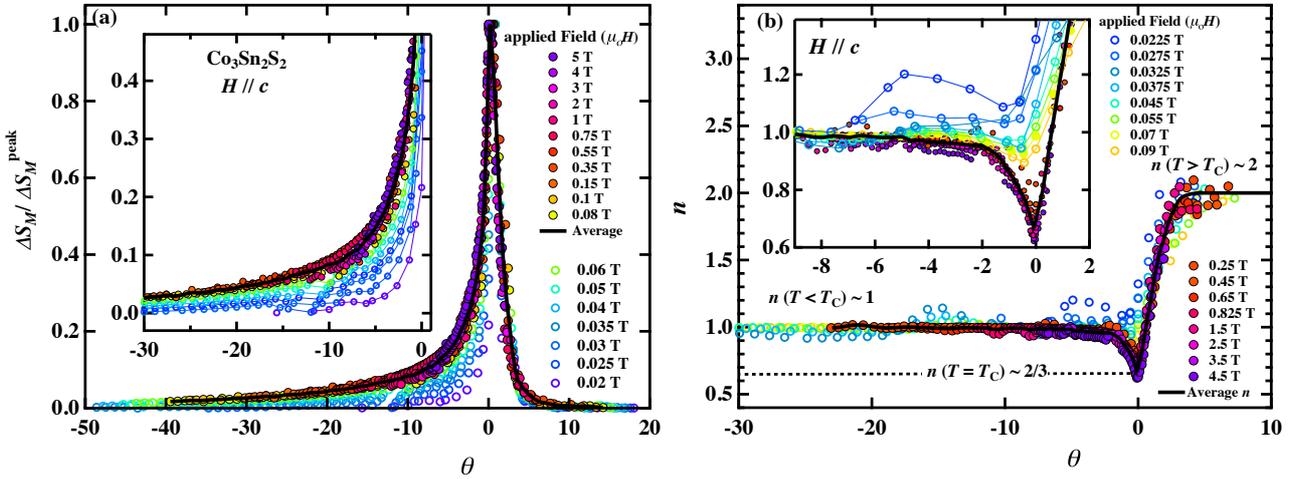

**Figure 5: Phenomenological scaling of the magnetic entropy change of Co$_3$Sn$_2$S$_2$:** (a) the normalized entropy change, $\Delta S_M(H,T)/\Delta S_M^{peak}(H,T)$, and (b) the generalized temperature- and field-dependent exponent $n(T,H)$, as functions of the reduced temperature, $\theta$. Curves at low fields show significant deviations from the universal curves of second-order magnetic transitions, as shown in a magnified portion in the inset.

As described in Appx. B, the generalized temperature- and field-dependent exponent $n(H_{int},T) = d\ln(|\Delta S_M(H_{int},T)|)/d\ln(H_{int})$ also has to fall in one universal curve when plotted against the reduced temperature, $\theta$[24,39]. Although the demagnetizing field slightly affects $\Delta S_M$, it has a dramatic influence on the field-exponent $n$[28]. We assure the estimation of $n(H_{int}, T)$ using only the equilibrium state magnetization data. Figure 5(b) shows the scaled plot of $n(H_{int}, \theta)$ for Co$_3$Sn$_2$S$_2$ using the same $T_{r1}(H_{int})$ and $T_{r2}(H_{int})$ of the scaling plot of $\left|\Delta S_M(H_{int},T)/\Delta S_M^{peak}(H_{int},T_C)\right|$. Similar to the scaling of $\left|\Delta S_M(H_{int},T)/\Delta S_M^{peak}(H_{int},T_C)\right|$, $n(H_{int}, \theta)$ correspond to fields above 0.1 T are well scaled to a universal curve in the whole temperature range, whereas the lower field $n(H_{int}, \theta)$ deviates from the universal scaling curve below $T_C$. As theoretically predicted for second-order MTs[24,40], the scaled $n(H_{int}, \theta)$ at high field exhibits a minimum, $\simeq 0.65$, at around $T_C$, $\simeq 177$ K, and approaches 1.0 far below $T_C$ and

2 far above $T_C$, which is observed here with low-field deviations only close to $T_C$ as illustrated in Fig. 4(b). The values of $n(H_{int}, \theta)$, 1 and 2 far below and above $T_C$, respectively, are the expected value in the mean field case, and are consistent with the linear behavior far below and above $T_C$ in the standard Arrot plot as shown in Fig. 3(b). We should note again that the $n(H_{int}, \theta)$ data interestingly show significant deviations from the universal curve, at and just below $T_C$, only at low internal fields ($\mu_0 H_{int} < 0.08 \sim 0.09$ T at $T = 162 - 177$ K), which is in consistence with the scaling results of $M(H_{int}, T)$ and $\Delta S_M(H_{int}, T)$.

Finally, the standard critical scaling of $\Delta S_M(H_{int}, T)$ around the transition temperature in a system with a second order MT[24,39], represented in Eq. B(7), had been attempted. Figure 6 is the scaling plot of $\Delta S_M(T, H_{int})(\mu_0 H_{int})^{\frac{\alpha-1}{\Delta}}$ vs. $\varepsilon(\mu_0 H_{int})^{-1/\Delta}$, which clearly confirms the violation of a second-order MT universality around and below $T_C$ only at low internal fields, $\mu_0 H_{int} \lesssim 0.1$ T.

4. **Discussion**

    In view of recent microscopic observations[21,41], we here discuss the macroscopic features of the controversial low-filed magnetic transition to an anomalous magnetic state in the ferromagnetic phase of the anisotropic kagomé ferromagnet $Co_3Sn_2S_2$[14,20,21,41,42], based on the present critical scaling analyses of the magnetization and MCE.

    Firstly, we summarize again the previously reported anomalous magnetic features observed in the macroscopic magnetization and AC susceptibility measurements[20] and their current physical interpretation based on the recent Lorentz TEM observation[21]. The abrupt increase at $T_C$ in the static magnetization[9,13,20,22], shown here in Fig. 2, as well as in the saturating magnetic field in the H-T phase diagram were observed[20], which are preliminary signatures of an unconventional MT at $T_C$. Further, the low-field anomalous magnetic state in the vicinity of $T_C$, named as "A-phase", with very slow spin dynamics of relaxation time of about 10 s was found[20]. This slow-dynamics anomalous magnetic state exists only within a zero- internal-field region due to the demagnetization effect, which imply the origin of this anomalous state is domain dynamics. Indeed, characteristic spontaneous magnetic domain motion and reorganization, a spontaneous bubble domain formation, were observed by the Lorentz TEM in the temperature- and field-region of the A-phase[21] and recently via MFM investigations[41]. With slow cooling through $T_C$, high-density domain bubbles has been observed just below $T_C$ only in low or zero magnetic field which should be attributed to high mobility of domain walls in the A-phase[20,21]. These observations imply us that the highly mobile domain dynamics governs the magnetic behaviors in large and long-time scales. Even in the finite internal-field region, the violation of the standard critical scaling below $T_C$ in very low internal field is a signature of these features. In rather high field or far from $T_C$, the ferromagnetic spin correlations in short scale govern the magnetic behaviors and the standard ferromagnetic critical behaviors are observed, whereas, in the low field region close to $T_C$, the

large-scaled domain dynamics become dominant and striking deviation from the standard ferromagnetic critical scaling is emergent.

Although the low magnetization of this compound that becomes lower close to $T_C$ has limited the contrast and prevented a spin configuration inside the domain bubbles in the Lorentz TEM study[21], the relatively large bubble size excludes the skyrmions textures, at least those stabilized by the Dzaloshinski-Moriya interaction or spin frustration. Even though, skyrmionic spin textures with relatively large size have been observed in uniaxial ferromagnets which are stabilized via the classical anisotropic dipolar interaction[43]. Our currently presented results of unconventional critical behaviors with scaling break down at low fields provide a further macroscopic evidence of an anomalous MT to an expected non-trivial magnetic state in the A-phase of $Co_3Sn_2S_2$ rather than a MT to a simple FM or AFM ordered state. These observed features may be common in systems with competing interactions. Further microscopic investigations are on demand to explore the spin textures just below $T_C$ for $Co_3Sn_2S_2$ and other uniaxial quasi-2D ferromagnets.

5. **Summary**

We have carefully studied the critical behaviors of the magnetization and magnetocaloric effect of $Co_3Sn_2S_2$ by using a high-quality single crystal. The analysis of numerous magnetization isotherms resulted in a modified Arrot plot and magnetic entropy changes around $T_C \simeq 177$ K with critical exponents, $\beta = 0.338(11)$, $\gamma = 1.137(23)$ and $\delta =$ 4.364(129), which do not satisfy predictions of the conventional theoretical models of second-order MTs. Scaling analyses of the magnetization $M(H_{int}, T)$, the magnetic entropy change $\Delta S_M(H_{int}, T)$ and the field-exponent $n(H_{int}, T)$ of $\Delta S_M$ show deviations from the universal curves in the ordered state only at low magnetic fields ($\mu_0 H_{int} \lesssim 0.1$ T) in the vicinity of $T_C$. The deviation from the standard second-order ferromagnetic critical scaling indicate an unconventional magnetic transition to an anomalous magnetic state of $Co_3Sn_2S_2$ that needs further microscopic explorations.

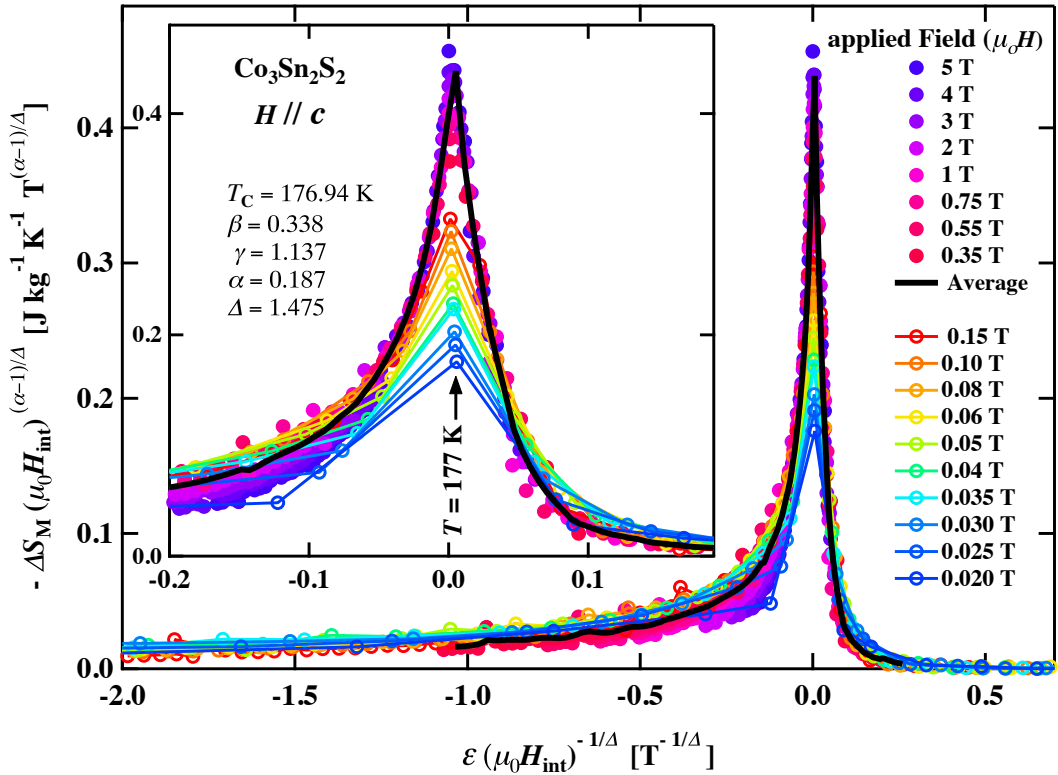

**Figure 6: Standard critical scaling of the magnetic entropy changes of $Co_3Sn_2S_2$:** scaling plot of the magnetic entropy change based on the Eq. (B7) in Appex. B. The inset shows a magnification around $T_C$.

**Acknowledgement**
M. A. K. would like to acknowledge a fellowship by the Ministry of Higher Education (Egypt) to Kyoto University, Japan.

**Appendix A: Background of the critical phenomenon study**

It is well known that in a second-order MT, the order parameter (i.e., magnetization *M*) is very small in the vicinity of the critical temperature as it changes continuously from zero. Thus, the Landau's expansion for the magnetic free energy at the ground state (that keeps a time-reversal symmetry in the absence of applied magnetic field *H*) and the corresponding magnetic-equilibrium equation of state with applied *H* are described, respectively, as[44]:

$$F_m(M) = F_m(0) + \frac{A}{2}M^2 + \frac{B}{4}M^4 + \cdots - HM, \tag{A1}$$

$$H = AM + BM^3, \tag{A2}$$

where *A* and *B* are temperature-dependent factors with *B* always positive while *A* changes its sign at the critical temperature to satisfy the zero and finite magnetizations respectively above and below the critical temperature. For a ferromagnetic system, *A* can have the form $A = a\varepsilon$ with $\varepsilon = (T - T_C)/T_C$ is the reduced temperature, *a* is a constant and $T_C$ is the Curie temperature. Thus, Eq. (A2) implies a parallel line of $M^2$ vs. $H/M$, the well-known Arrot plot[45], with zero abscissa and ordinate intersections for the curve at $T = T_C$.

A more general equation of the magnetic state that describes the isothermal magnetization $M(H_{int})$ around a second order MT is the Arrot-Noaks equation of state[46]:

$$\left(\frac{H}{M}\right)^{\frac{1}{\gamma}} = a\,\varepsilon + B\,M^{\frac{1}{\beta}}, \tag{A3}$$

where $\beta$ and $\gamma$ are two of the critical exponents and are related to a third critical exponent $\delta$ by the Widom scaling relation $\delta = 1 + \frac{\gamma}{\beta}$[47]. The critical exponents $\beta$, $\gamma$ and $\delta$ are respectively associated with the spontaneous magnetization $M_s(T) = \lim_{H \to 0} M$, the inverse initial susceptibility $\chi_0^{-1}(T) = \lim_{H \to 0} \frac{H}{M}$ and the critical magnetization isotherm $M(H, T_C)$ by the power laws[30],

$$M_S(T) = M_0 (-\varepsilon)^\beta, \quad T < T_C \ (\varepsilon \text{ and } A > 0), \tag{A4}$$

$$\chi_0^{-1}(T) = \left(\frac{H}{M_0}\right) \varepsilon^\gamma, \ T > T_C \ (\varepsilon \text{ and } A < 0), \tag{A5}$$

$$M = \Gamma H^{\frac{1}{\delta}}, \quad T = T_C \ (\varepsilon \text{ and } A = 0), \tag{A6}$$

where $M_0$, $H/M_0$ and $\Gamma$ are the critical amplitudes of corresponding quantities. Above-mentioned equation of state (Eq. (A2)) is a special case with $\beta = 1/2$, $\gamma = 1$ and $\delta = 3$ (the mean-field criticality). Critical exponents of some representative models, the 3D Heisenberg, 3D Ising, 3D XY, tricritical mean-field models are listed in table 1.

Eqs. (A4) and (A5) can be re-formed to:

$$M_S(T)[dM_S(T)/dT]^{-1} = (T - T_C)\beta^{-1}, \text{ for } T < T_C, \tag{A7}$$

$$\chi_0^{-1}(T)[d\chi_0^{-1}(T)/dT]^{-1} = (T - T_C)\gamma^{-1}, \text{ for } T > T_C. \tag{A8}$$

Thus, the critical exponents $\beta$ and $\gamma$ and the MT temperature $T_C$ can be estimated from the $M_S(T)[dM_S(T)/dT]^{-1}$ vs. $T$ and $\chi_0^{-1}(T)[d\chi_0^{-1}(T)/dT]^{-1}$ vs. $T$ plots, respectively (Kouvel-Fisher method[30]).

Based on the scaling hypothesis of the second order phase transition[48], the asymptotic equation of state in the vicinity of the MT temperature, the scaling form of the magnetization isotherms, is given as,

$$m = f_\pm(h), \tag{A9}$$

where,

$$m = |\varepsilon|^{-\beta} M(H, \varepsilon) \text{ and } h = H|\varepsilon|^{-(\beta+\gamma)}. \tag{A10}$$

The Arrot-Noaks equation of state, Eq. (A3), is a useful phenomenological expression of the scaling equation (Eq. (A9)). By introducing the correct values of $\beta$ and $\gamma$ to Eq. (A9), the plots of $m$ as a function of $h$ around $T_C$ should fall onto two universal curves, one with the regular function $f_+$ for data above $T_C$ and another with $f_-$ for data below $T_C$.

In systems exhibiting an unconventional magnetic order such as chiral helimagnets[49], frustrated magnets[50] and uniaxially anisotropic van der Waals ferromagnets[51] the order parameter would be not simply the magnetization $M$ but rather a multi-component. For instance, it can be described by a slowly varying periodic spin density in the B20 chiral helimagnets[49]. In these magnets, a metamagnetic transition drives the system from a low-field unconventional magnetic phase(s) to a field-forced ferromagnetic (FFM) state. Thus, below a critical magnetic field, these magnets usually violate the second-order MT. However, the MT thermally driven from the paramagnetic (PM) to FFM state above this critical field can be described by Eq. (A3) as a second-order MT with the order parameter $M$.

**Appendix B: Magnetocaloric effect (MCE)**

The change in entropy with applying magnetic field is related to the change in $M$ with temperature by the thermodynamic Maxwell relation,

$$\left(\frac{\partial S_M(T,H)}{\partial H}\right)_T = \left(\frac{\partial M(T,H)}{\partial T}\right)_H, \tag{B1}$$

i.e., the entropy change by magnetic field $\Delta S_M(T, H) = S(T, H) - S(T, 0)$ is given by,

$$\Delta S_M(T, H) = \int_0^H \left(\frac{\partial M(T,H)}{\partial T}\right)_H dH. \tag{B2}$$

In the real measurement of $M$ isotherms at small steps of discrete fields, $\Delta H$, and with small temperature intervals, $T_2 - T_1$, $\Delta S_M$ can be approximated as[31],

$$\Delta S_M\left(\frac{T_1+T_2}{2}, H\right) = \sum \left[\frac{M(T_2, H) - M(T_1, H)}{T_2 - T_1}\right] \Delta H. \tag{B3}$$

As the application of a magnetic field usually results in further order of the magnetic moments against thermal fluctuations, i.e., a decrease in the magnetic entropy, we observe negative $\Delta S_M$ in a conventional MCE. However, application of a magnetic field may result in $\Delta S_M > 0$, that indicates a magnetic field-induced disorder with respect to the magnetic ground state, i.e., in absence of the magnetic field. Typically, the ferromagnetic and antiferromagnetic ordered states correspond to the former (negative $\Delta S_M$) and the latter (positive $\Delta S_M$) cases, respectively.

In FM systems of second-order MT, the absolute value of the magnetic entropy change exhibits a peak at $T_C$ and the peak value, $\left|\Delta S_M^{peak}(H, T_C)\right|$, usually occurs at $T_C$ and increase with the applied field in power laws as[24],

$$\begin{cases} \left|\Delta S_M^{peak}(H, T_C)\right| \sim H^{n(T_C)}, \\ \delta T_{FWHM}(H, T_C) \sim H^b, \\ RCP \sim H^c, \end{cases} \tag{B4}$$

The exponents $n(T_C)$, $b$, and $c$ are related with the critical exponents of magnetization $\beta$, $\gamma$, and $\delta$ as,

$$\begin{cases} n(T_C) = \frac{1-\alpha}{\Delta}, \\ b = \frac{1}{\Delta}, \\ c = n(T_C) + b = 1 + \frac{1}{\delta}, \end{cases} \tag{B5}$$

where $\alpha = 2 - 2\beta - \gamma$ and $\Delta = \beta\delta = \beta + \gamma$. This relation (B5) can be obtained by substituting the differential of the Arrott-Noakes equation of state, Eq. (A3), $\left(\frac{\partial M(T,H)}{\partial T}\right)_H$ into Eq. (B2)[38,39].

Further, a phenomenological universal scaling of the MCE was proposed[24,39]. In this scaling procedure, the normalized magnetic entropy change $\left|\Delta S_M(H,T)/\Delta S_M^{peak}(H, T_C)\right|$ and the temperature- and field-dependent exponent $n(H,T) = dln(|\Delta S_M(H,T)|)/dln(H)$ are plotted against a reduced temperature, $\theta$, which is obtained by imposing two reference temperatures $T_{r1}$ and $T_{r2}$ correspond to $\theta = \pm 1$ so that[38,39],

$$\theta = \begin{cases} -\frac{T-T_C}{T_{r1}-T_C}, & T \leq T_C \\ \frac{T-T_C}{T_{r2}-T_C}, & T > T_C. \end{cases} \tag{B6}$$

In spite of lack of the theoretical foundation, this phenomenological scaling has been confirmed to successfully describe the simulation data based on the mean field criticality and many experimental data[24,39]. Theoretically[39], a standard critical scaling form of $\Delta S_M(H, T)$ can be derived by introducing the scaling form of $M(H, T)$, Eq. (A9), into Eq. (B2) as,

$$\Delta S_M(T,H) = a_M |\varepsilon|^{1-\alpha} f(\frac{\varepsilon}{H^{1/\Delta}}) = a_M H^{\frac{1-\alpha}{\Delta}} g(\frac{\varepsilon}{H^{1/\Delta}}), \tag{B7}$$

where, $a_M$ is a constant and $g(x) = |x|^{1-\alpha} f(x)$. In consistence with Eqs. (B5), this equation, (B7), implies that if the magnetic entropy change is rescaled by a factor $H^{n(T_C)}$ and the reduced temperature by a factor of $H^b$, the experimental data should collapse onto the same curve[39].

**Tables:**

**Table 1:** Comparison of critical exponents of $Co_3Sn_2S_2$ to those proposed by various models.

| **Experiment/Model** | **Method** | **$\beta$** | **$\gamma$** | **$\delta = 1+ \gamma/\beta$** | **$T_C$ (K)$^*$** |
|---|---|---|---|---|---|
| $Co_3Sn_2S_2$ (this work) | modified Arrot plot | 0.343(2) | 1.116(15) | 4.254(48) | 177.13 |
|  | Kouvel-Fisher** | 0.338(11) | 1.137(23) | 4.364(129) | 176.94 |
|  | critical $M(H_{int})$† |  |  | 4.441(19) | 177 |
|  | MCE‡ | 0.3928 | 1.3878 | 4.7103 | 177 |
| Mean field theory | theory | 0.5 | 1.0 | 3.0 |  |
| 3D Heisenberg | theory | 0.365 | 1.386 | 4.797 |  |
| 3D Ising | theory | 0.325 | 1.241 | 4.818 |  |
| 3D XY | theory | 0.345 | 1.316 | 4.815 |  |
| Tricritical mean-field | theory | 0.25 | 1.0 | 5.0 |  |

* the Curie temperature values are estimated from data above the transition temperature,

** these values are selected to estimate the standard critical scaling of magnetization (Fig. 3d) and MCE (Fig. 6),

† value of $\delta$ obtained by fitting the critical magnetization isotherm $M(H, T_C)$ to Eq. (A6),

‡ critical exponents obtained from the magnetocaloric effect by fittings to Eqs. (B4) and using relations in Eqs. (B5).
∗ Author to whom correspondence should be addressed:
E-mail: makassem@aun.edu.eg